# Creating Nanostructured Superconductors On Demand by Local Current Annealing

Hongwoo Baek[1, 2,†], Jeonghoon Ha[1, 3,†], Duming Zhang[1, 3], Bharath Natarajan[1,3], Jonathan P. Winterstein[1], Renu Sharma[1], Rongwei Hu[4], Kefeng Wang[4], Steven Ziemak[4], Johnpierre Paglione[4], Young Kuk[2], Nikolai B. Zhitenev[1], and Joseph A. Stroscio[1,*]

[1]Center for Nanoscale Science and Technology, National Institute of Standards and Technology, Gaithersburg, MD 20899, USA
[2]Department of Physics and Astronomy, Seoul National University, Seoul, Korea
[3]Maryland NanoCenter, University of Maryland, College Park, MD 20742, USA
[4]Center for Nanophysics and Advanced Materials, Department of Physics, University of Maryland, College Park, MD 20742, USA
†These authors contributed equally to this work.

**Abstract:**

Superconductivity results from a Bose condensate of Cooper-paired electrons with a macroscopic quantum wavefunction. Dramatic effects can occur when the region of the condensate is shaped and confined to the nanometer scale. Recent progress in nanostructured superconductors has revealed a route to topological superconductivity, with possible applications in quantum computing. However, challenges remain in controlling the shape and size of specific superconducting materials. Here, we report a new method to create nanostructured superconductors by partial crystallization of the half-Heusler material, YPtBi. Superconducting islands, with diameters in the range of 100 nm, were reproducibly created by local current annealing of disordered YPtBi in the tunneling junction of a scanning tunneling microscope (STM). We characterize the superconducting island properties by scanning tunneling spectroscopic measurements to determine the gap energy, critical temperature and field, coherence length, and vortex formations. These results show unique properties of a confined superconductor and demonstrate that this new method holds promise to create tailored superconductors for a wide variety of nanometer scale applications.

---

[*] To whom correspondence should be addressed: joseph.stroscio@nist.gov

# I. INTRODUCTION

Superconductivity is characterized by specific characteristic length scales: the coherence length $\xi$, the length scale of the Cooper-paired electrons, and the London penetration length $\lambda$, the length scale of magnetic field penetration into the superconductor [1]. When the size of a superconductor becomes comparable with these characteristic lengths, the properties of the superconductor can change dramatically, including the critical temperature, critical magnetic field, and vortex structures. Early work focused on the effect of condensate confinement in mesoscopic disks [2–4], and the effect of symmetry [5] in comparing disks, squares [6], and mesoscopic triangles [7]. Focus on vortex states in mesoscopic superconductors, both theoretical and experimental, revealed multi-vortex Abrikosov-like states with spatial arrangements of singly quantized vortices, or giant vortex states depending on details of the condensate confinement [8–16].

More recently, interest in nanostructured superconductivity as a route to topological superconductivity has emerged [17]. Topological superconductors are analogs of topological insulators. Topological superconductors are composed of a superconducting gap in the bulk and delocalized Andreev surface states at zero energy, which correspond to the so-called "zero modes" of Majorana fermions [18,19]. Initial proposals for topological superconductors focused on combining topological insulators in proximity with a conventional superconductor [20]. Experimental demonstration of proximity induced superconductivity in a topological insulator has been observed using transport measurements [21,22] and scanning tunneling spectroscopy [23,24]. Rare-earth half-Heusler compounds, which have a crystal structure that lacks inversion symmetry, have been predicted to be topologically non-trivial depending on the chemical composition [25,26]. Due to both topological insulator and superconducting properties, half-Heusler materials have been predicted to offer multifunctional topological devices [27], depending on lattice parameters and spin-orbit coupling strengths. YPtBi was recently shown to have superconducting properties with low carrier concentration in electron transport measurements, with a critical temperature of 0.77 K and an upper critical magnetic field of 1.5 T [28]. Band calculations of YPtBi show a single band inversion at the $\Gamma$ point, and therefore it is predicted to be a topological material [29,30]. Based on a non-centrosymmetric and strong spin-orbit coupling, YPtBi is considered a candidate for topological superconductivity with triplet pairing [28].

In this article, we report scanning tunneling microscopy (STM) measurements on superconducting nanostructured disks of polycrystalline YPtBi, created by local current annealing inside the STM tunneling junction. The superconducting properties of these nanostructures were obtained from tunneling spectroscopy measurements as a function of spatial position, temperature, and magnetic field. The spectra show non-Bardeen-Cooper-Schrieffer (BCS) behavior, possibly suggesting non-conventional pairing, as proposed by previous transport measurements [28]. Different energy gaps, critical temperature, and critical magnetic fields were found in the center region versus the perimeter region of the nanostructure. Spatial measurements determined the coherence length to be ≈15 nm. Magnetic field measurements show the sequential addition of single vortices to the nanostructure, cumulating into a giant vortex ring at 1.25 T. These measurements demonstrate a novel method of creating tailored nanostructured superconductors with complex superconducting materials for a variety of future applications.

## II. CREATING SUPERCONDUCTING NANOSTRUCTURES IN A DISORDERED YPtBi SURFACE

The nanostructured superconductors were made from single YPtBi crystals. The bulk crystals were grown in excess molten bismuth [31]. The crystal has cubic structure with a lattice constant of 0.665 nm [Fig. 1(a)]. A bulk crystal sample was cleaved in ultrahigh vacuum in a base pressure of $5 \times 10^{-9}$ Pa, and transferred *in situ* to a cryogenic STM system with a base temperature of 10 mK [32]. Topography on the cleaved surface showed a disordered cleaved surface with a granular structure with a height variation of ≈ 1 nm (see Fig. A2). Atomic resolution was observed on some of the grains. The structure on the surface is made from an amorphous region of the bulk crystal (see TEM results in Appendix C), which determined the cleaved surface during the cracking of the crystal. We created a local nanoscale superconducting disk by crystallizing the surface structure under the tip of the STM junction [Fig. 1(b)]. The method uses a pulsed current of 10 ms duration through the tunneling junction by applying biases of 6 V to 8 V, while the tip is held fixed at the nominal tunnel junction distance. The current pulse achieved a local structural transition to a superconducting phase, forming circular islands in topographic images [Fig. 1(c)]. A variety of islands could be created using this method by positioning the STM probe to different regions of the surface, as shown in Fig. 1(d). The islands have dimensions on the order of 100 nm in diameter. We focus on the island [Fig. 1(c)] in the red box in the lower left of Fig. 1(d) to examine its superconducting properties.

Tunneling spectroscopy is a useful probe of the density of states of the normal and superconducting properties of a material. In Fig. 2(a) we observe that the density of states of the nanostructured island and pristine cleaved surface appear to be similar. This reflects the similar chemical compositions in the regions as determined by the TEM measurements discussed below.

The superconducting properties of the island are revealed as we reduce the energy range of the measurements revealing a superconducting gap, as shown in Fig. 2(b). Surprisingly, no superconductivity is observed in the tunneling spectroscopy measurements from the as-cleaved disordered surface, as we compare spectra off and on the island [Fig. 2(b)], even though the bulk crystal is superconducting as verified by bulk transport measurements (see Appendix D). Instead, the spectrum in Fig. 2(b) off the island shows metallic states around the Fermi energy (zero bias) without a superconducting gap. Focusing on a further smaller energy range, the spectra in Fig. 2, (c) and (d) on the island clearly shows a well-defined superconducting gap and symmetric coherence peaks in the bias voltage range of ± 1 mV. The island perimeter appears to have a smaller superconducting gap compared with the center of the island, as observed by comparing Figs. 2, (c) and (d), and the gaps determined by the fits to the modified BCS theory by Maki [33,34] are shown by the solid lines.

In order to clarify how we created a superconductive phase in the nanostructures on a disordered surface, we intentionally created 7 superconducting islands with 1 μm spacing [Fig. 1(d)] and analyzed their lattice structure and chemical composition [Fig. 1, (e) and (f)] using transmission electron microscopy (TEM). Depending on the DC bias level of the pulse and junction impedance (tip-sample distance), the shape and the size of the nanostructures could be varied, as shown in Fig. 1(d). Electron transparent cross-sections of the nanostructures were vertically sliced out using a focused ion beam (FIB) microscope for TEM measurements, which allow a comparison of the differences between nanostructures, bulk and pristine surface regions (see Appendix C). In the cross-sectional TEM images, we found an amorphous surface region with a thickness of ≈ 50 nm, nanostructures imbedded in the amorphous region, and the underlying half-Heusler lattice in the bulk region [Fig. 1(e)]. The 50 nm amorphous layer is thick enough to screen proximity of superconductivity from the bulk crystal below, as seen from the coherence length of the material that we determined below. X-ray energy-dispersive spectroscopy (XEDS) determined a chemical composition containing yttrium, bismuth, and platinum, which is expected from the bulk chemical composition, and additional oxygen in surface regions, presumably from exposure to the atmosphere. The superconducting nanostructures which were created in the amorphous layer show a 1:1:1 atomic ratio of yttrium, bismuth and platinum as expected for the YPtBi half-Heusler compound. In high resolution TEM images, the bulk region shows single crystalline structure [Fig. 1(f)] with the lattice constant matched to the half-Heusler structure. In the nanostructure, small crystallites were observed with lattice spacings close to the bulk YPtBi. The crystallites show up as a ring pattern in diffraction images and fast-Fourier transforms of high resolution images due to the orientational degree of freedom of the polycrystalline particles (see Fig. A5). The fractional crystallization to the half-Heusler structure could cause a phase transition to a superconducting phase from the metallic amorphous region, which is locally created by current annealing through the STM probe.

## III. PROXIMITY EFFECT ACROSS THE SUPERCONDUCTING NANOSTRUCTURE

When a superconductor (S) is in electrical contact with a normal metal (N), the "leakage" of Cooper pairs causes the normal metal to acquire superconducting-like properties. This phenomena, known as the *proximity* effect, has been of renewed interest due to the pursuit of obtaining a topological superconductor by combining conventional superconductors with topological insulators [20–24]. Additional measurement opportunities have resulted from the ability to microfabricate mesoscopic structures combined with the ability to probe the proximity effect with nanometer spatial resolution using scanning tunneling spectroscopy [35–38]. At a conventional NS junction, superconductivity is suppressed and exponentially decays in the normal metal region following the nonlinear Usadel equation of the superconducting order parameter [39]. As schematically shown in Fig. 3(c), the order parameter will decay in the normal metal with a coherence length, $\xi_N$, which depends on the mean free path of the metal near the junction, and will decay at the superconductor boundary with coherence length, $\xi_S$.

To examine the proximity effect at the NS boundary of the nanostructure, we mapped the superconducting properties across the superconducting islands by measuring the differential tunneling conductance as a function of spatial position [Fig. 3]. Figure 3(b) shows the spatial dependence of tunneling conductance across the vertical line in Fig. 3(a). The blue region shows a well-defined superconducting gap within the island, which becomes gradually filled with intensity at the NS boundary. We use the zero bias conductance (ZBC) as a measure of the superconductor order parameter. Note a low value of ZBC indicates a high value for the gap and hence the order parameter, therefore we plot the ZBC with increasing values in the downward direction. Figure 3(e) shows the ZBC as function of spatial position across the island following the line in Fig. 3(a). The superconducting order parameter starts out at a maximum in the center of the island, and decays slowly to the island boundary and then rapidly decays in the normal region surrounding the island. Figure 3, (d) and (f) show a close up of the exponential decay into the normal metal regions on the sides of the islands. An exponential fit yields a normal coherence length of $\xi_N = (20 \pm 2)$ nm [40], which being less than 50 nm amorphous layer thickness, explains the lack of superconductivity away from the annealed islands. A similar value for the superconductor coherence length is inferred by the length scale the ZBC decays in the interior of the island [Fig. 3(e)], comparing favorably to the value of 15 nm obtained from upper critical field measurements [28].

## IV. SUPERCONDUCTING PROPERTIES OF NANOSTRUCTURED YPtBi ISLANDS

The superconducting properties of a material are reflected in the pairing energy, coherence length, and critical temperature and magnetic field, which we examine in detail in this section for the nanostructured island. Non-conventional behavior is seen in deviation from BCS theory. To determine the energy gap we fit the spectra to the Maki theory, which is an extension of the BCS theory, accounting for effects of orbital depairing, the Zeeman splitting of the spin states, and spin-orbit scattering [33,34]. The fitting parameters for zero field are the energy gap, $\Delta$, the orbital depairing parameter, $\zeta$, and the effective temperature, $T_{\text{eff}}$. A detailed fit of the spectra reveals an inhomogeneity of the gap size observed from the center to perimeter regions of the island, with $\Delta = (220 \pm 1)$ μV measured on the center [Fig. 2(c)] and $\Delta = (190 \pm 1)$ μV on the perimeter of the island [Fig. 2(d)] [40]. Additionally, we note that the tunneling spectra in Fig. 2, (c) and (d) are not fully gapped and contain finite mid-gap states in the superconducting gap for both the center and perimeter regions. These mid-gap states might be related to a topologically nontrivial pairing mechanism [41], or possibly to the polycrystalline nature of the island structure [42].

To determine the critical values of the superconducting nanostructure, we have measured a series of tunneling spectra with increasing temperature and magnetic field, and obtained detailed fits of the energy gap of the nanostructured island [Fig. 4]. The superconducting gap and coherence peaks are fully suppressed at the temperature of 1.4 K, as observed in Fig. 4(a). Based on fits to BCS theory [1], a critical temperature of $T_C = (1.37 \pm 02)$ K was obtained.[Fig. 4(b)] [40]. In Appendix D we show the measured bulk resistance from the same crystals, which show a superconducting critical temperature of ≈ 0.75 K, in agreement with previous measurements [28]. The higher critical temperature of the nanostructured island may result from its small size or multicrystalline structure.

Figure 4, (c)and (d) show the effect of increasing magnetic field on the tunneling spectra and gap. With increasing magnetic field the superconducting gap decreased and went to the normal state at 2.25 T. This value thus corresponds to the upper critical magnetic fields, $B_{C2}$, for the nanostructured island. Measurements of the critical field as a function of temperature for the bulk crystal are shown in Appendix D.

## V. SEQUENTIAL ADDITION OF SINGLE VORTICES INTO A NANOSTRUCTURED ISLAND

At magnetic fields between the lower critical field, $B_{C1}$, and the upper critical field, $B_{C2}$, partial penetration of magnetic field is allowed in a type II superconductor and the penetration can be observed in the form of vortex formation. To observe the vortex formation within the superconducting island, we carried out a spatial mapping of the differential conductance and extracted maps at zero bias as a function of magnetic field, as shown in Fig. 5. Red represents the superconducting region with low conductance in the gap and blue corresponds to the normal state. With increasing magnetic field, the conductance maps show a single vortex sequentially added to the superconducting island starting from a lower critical field of $B_{C1} = 0.2$ T. This value for the lower critical field is much higher than previous bulk measurements of 0.008 mT, due to the confined nanostructure [43]. The vortices occupy the perimeter region with equal spacing. The vortex structures remain fixed and stable during the acquisition time of the measurement, which can be many hours long. Superconductivity in the center region is gradually suppressed without any vortex structure appearing. In contrast, in the perimeter region, a single vortex is added every 0.0625 T between 0.2 T and 0.5 T, and then the change in magnetic field required for creating single vortex increases to the twice this number, 0.125 T, at $B = 0.5$ T.

The magnetic flux per vortex can be related to the parity of the superconductor. In the case of chiral *p*-wave superconductors, a first order transition between single- and double-winding vortex states is theoretically expected at an intermediate magnetic field, $B_D$ [44]. For magnetic fields of $B_{C1} < B_D < B < B_{C2}$, a flux lattice of doubly quantized vortices is supposed to be stable. A doubling of the magnetic field per single vortex, as observed in the results shown in Fig. 5, is possible due to the double winding vortex states of a *p*-wave superconductor [44]. It is instructive to plot the vorticity vs. normalized flux to examine the vorticity behavior [45]. The normalized flux is the magnetic flux, $\Phi_d = B\pi(D_d/2)^2$, normalized by the flux quantum $\Phi_0 = 2.067 \times 10^{-15}$ Tm$^2$, where $D_d$ is the island diameter. As seen in Fig. 6 the vorticity is below the solid line, which is the case for $L = \Phi_d/\Phi_0$. The difference of the vorticity from the solid line is due to the diamagnetic response of the island and the presence of large screening currents, and has been observed in previous measurements on larger superconducting disks [16,45]. What is unique in the present measurements is the kink in slope at *L*=4. This may result from vortex-vortex interactions, which will depend on the particular vortex configuration, and are altered at different fields. Confinement effects due to the small size of the nanostructure compared to the coherence length may also play a role. The average radius of the vortices in Fig. 5 is ≈14.5 nm, which is close to the coherence length obtained from the upper critical field measurements.

Quite interestingly, the vortices do not penetrate the center of the island. Instead, as the vortices are added with increasing field in Fig. 5, we observed a one dimensional vortex distribution on an annulus, and found that the vortex spacing (*a*) did not follow the Abrikosov lattice derived from the Ginzburg-Landau equation, $a = (2\Phi_0/\sqrt{3}B)^{1/2}$ for a close-packed hexagonal array; this equation leads to sparse vortices at low fields ($H < 0.7$ T) and dense vortices at high fields ($H > 0.7$ T). In previous measurements of mesoscopic superconducting disks [4,16,45], the first vortex is typically observed in the center of the disk in contrast to the results in Fig. 5. These previous measurements were on disks with diameters in tens of micrometers, much larger than the coherence length. In contrast, our islands are two orders of magnitude smaller, with a size on the order of ten coherence lengths. When the diameter is only a few coherence lengths, STM tunneling measurements on Pb islands have observed only a few vortices or a single giant vortex state [9,11,13]. Our island is somewhere in between these length scales and possesses an array of single vortices at low fields, which appear to fuse into a giant vortex annulus at 1.125 T.

The formation of giant vortex states typically occurs in the center of a superconducting island [9–11,13], and requires the following criteria: the lateral size of the nanostructure should be several times the coherence length for a strong confinement effect, and the upper critical field of the superconducting material should be high enough to observe the merging of vortices. In our system the vortices occupying an annulus around the center region, with the lack of a vortex in the center. We can examine the confinement for our vortices in two ways. Taking $D_d$ from the measured diameter of the island, we have $D_d \approx 12\xi$, which is in the strong confinement regime. On the other hand if we take $D_d$ to be the thickness of the annulus, we have $D_d \approx 3\xi$, which resembles a thin narrow superconductor that should have vortices arranged in a single row when the width $w \lesssim 6\,\xi$. In the creation of our nanostructured island, it is likely that the current pulse would have a spatial profile, which decays with distance from the center, leading to an inhomogeneity in the island. For example, the center of the island might be more crystalline than the perimeter, and thus forming a barrier or defect in the center. Numerical calculations have shown that such barriers or defects can lead to complex vortex patterns [46–49]. Another possibility for the non-conventional vortex structure observed in Fig. 5 could be due to the possibility of multi-band superconductivity, which can also lead to non-conventional vortex structures [50–54]. As the characteristics of the superconducting state in YPtBi are still largely unknown, further experimental and theoretical work in the interesting material is needed to fully explain these observations.

## VI. CONCLUSIONS

In conclusion, we report a reproducible method to create superconducting YPtBi nanostructures by local current annealing of structurally amorphous YPtBi in an STM tunnel junction. We observed non-BCS tunneling spectra and strong confinement effects with novel vortex structures. Looking toward the future, we anticipate that having the ability to induce a local structural transition to a superconducting phase with local probes will be a useful quantum nanotechnology workbench to design superconducting nanostructures for investigating both fundamental interest and practical applications of quantum electronics. We envision the ability to create networks of low dimensional superconducting structures, such as lines and dots, or create patterns to engineer designer bosonic systems with artificial gauge fields [55,56].

## ACKNOWLEDGEMENTS

J.H., and D.Z. acknowledge support under the Cooperative Research Agreement between the University of Maryland and the National Institute of Standards and Technology Center for Nanoscale Science and Technology, Grant No. 70NANB10H193, through the University of Maryland. H.B. and Y.K. are partly supported by Korea Research Foundation through Grant No. KRF-2010-00349. Crystal growth and characterization at the University of Maryland were supported by a DOE-Early Career Award (Grant No. DE-SC-0010605). We would like to thank Guru Khalsa for valuable discussions, Steve Blankenship, Glen Holland, and Alan Band for technical assistance.

**APPENDIX A: SURFACE PREPARATION AND STM IMAGING OF CLEAVED YPTBI**

The YPtBi crystal was cleaved at room temperature in ultra-high vacuum (UHV) and then transferred into the STM module [32]. In the STM module, the tip was aligned onto a terrace of the cleaved surface, as shown in the optical image in Fig. A1. Following alignment, the STM module is lowered into a dilution refrigerator and cooled to 10 mK.

STM topography measurements on the cleaved surface are shown in Fig. A2. Figure A2(a) shows the roughness of the cleaved surface over 50 nm x 50 nm with a height scale of 1.7 nm. Fig. A2(b) shows zoomed in views with image sizes of 20 nm x 20 nm, and 5 nm x 5 nm in the inset. The smaller image in the inset shows atomic resolution features on the small clusters.

**APPENDIX B: TUNNELING SPECTROSCOPY MEASUREMENTS**

The tunneling spectroscopy measurements were made at a base temperature of ≈10 mK. Electrical noise limits the effective temperature to values higher than the system base temperature, which is determined by fitting the spectra of a known superconductor. We use thin Al films grown in-situ to determine the effective noise temperature of our system. Figure A3 shows the tunneling spectrum of an Al thin film (blue circles), which is fully gapped. The red line shows a calculation using the modified BCS Maki theory [33,34], with an effective temperature of $T_{\text{eff}} = (160\pm5)$ mK [40]. Comparing the Al spectrum with the spectrum from the YPtBi nanostructured island in Fig. 2(c), we observe that the finite gap structure in the YPtBi superconductor is not instrumentally limited, as evidenced by the ability to fully resolve the Al gap in Fig. A3.

## APPENDIX C: TEM MEASUREMENTS

The crystal structure of the YPtBi nanostructures were examined with TEM. Figure A4 describes the sample preparation using focused ion beam milling to cut a wedge shaped sample, and then thinned down for TEM cross sectional measurements. Figure A5 summarizes the TEM measurements. Figure A5(b) shows the TEM cross section image from the boundary region between the amorphous top layer and underlying crystal, outlined by the red box in Fig. A5(a). The underlying crystal structure is in agreement with the crystal structure for YPtBi, while the amorphous region shows no well-defined structure. Figure A5(c) shows the cross sectional image of the area within the superconducting nanostructure. Here small crystalline grains can be observed, which give rise to rings of spots in the fast-Fourier-transform of this image, shown in Fig. A5(d).

## APPENDIX D: ELECTRICAL TRANSPORT MEASUREMENTS

Electrical transport measurements were carried out on a sample from the same crystal batch used for STM measurements, which are summarized in Fig. A6. The sample shows superconducting characteristics similar to previous measurements [28], with the exception of a broad decrease in resistance preceding the sharp superconducting transition at around ≈0.75 K.

This may be due the indium electrodes becoming superconducting at low temperature. This broad transition was suppressed as indium becomes normal under a moderate field [Fig. A6(c)]. The upper critical field measurements in Fig. A6(c) are also in good agreement with previous measurements. The main result of these measurements confirms that the same sample used for STM measurements was superconducting, with nominally the same characteristics as bulk YPtBi.

The uncertainty was compounded from standard deviation values of slope in linear fits to the baseline and sidewall [40].

**Figure Captions**

FIG. 1. (color online) Creating YPtBi superconducting nanostructured islands by local current annealing. (a) C1b crystal structure of YPtBi: Y (purple), Pt (grey), Bi (green). (b) Illustration of the process of creating superconducting islands by passing a current pulse with the STM probe tip held fixed within ≈1 nm of an amorphous YPtBi layer. (c) 3D rendered STM topographic image (200 nm x 200 nm) of a created superconducting island (see red box in (d)). (d) SEM image showing seven different superconducting nanostructures created by local current annealing with the STM probe tip. The island in the lower left corner, outlined by the red box, is used for the superconducting property measurements shown in Figs. 2-6. (e) TEM cross section of the superconducting island in the bottom center of (d) highlighted by the orange line. The cross sectional image shows ≈50 nm thick amorphous layer covering the entire surface of the YPtBi crystal. The white arrow indicates the boundary between the amorphous layer and underlying crystal. The superconducting nanostructure created by local current annealing is outlined by the white oval. Most of the nanostructures are characterized by a dimple in the center of the structure. (f) High resolution TEM image of the boundary region separating the underlying YPtBi crystal and the region of the created nanostructure. TEM analysis shows that the superconducting nanostructure contains more crystalline domains than the surrounding amorphous regions, which were not subject to local current annealing.

FIG. 2. (color online)Tunneling spectroscopy of a superconducting nanostructured island. (a) Comparison of tunneling spectra on the pristine cleaved surface and nanostructured island for a large sample bias range of ±100 mV. Over this range of tunneling bias the pristine cleaved surface and nanostructure appear similar. (b) With a smaller bias range of ±10 mV, the nanostructured island displays a superconducting gap at very low bias. (c,d) Differential tunneling conductance spectra (blue dots) over ±1mV sample bias measured over the center and perimeter of the nanostructured island in Fig. 1(c), respectively. Note in both cases the spectra are not fully gapped and show some conductance within the gap. The solid red lines are fits to the modified BCS Maki theory [33,34] yielding the following parameters: island center, $\Delta_0$=(222±1) μeV, $\zeta$=0.019 ± 0.001, $T_{eff}$=150 mK, and island perimeter, $\Delta_0$=(198±1) μeV, $\zeta$=0.029 ± 0.002, $T_{eff}$=150 mK [40].

FIG. 3. (color online) Proximity effect across a nanostructured superconducting island. (a) STM topographic image, 190 nm x 300 nm, of the superconducting island. The height scale from dark to bright is 9.5 nm. (b) dI/dV line scan as a function of position through the center of the island along the green line in a. The *dI/dV* is normalized and shown in a color scale. The blue area denotes larger gap values, $\Delta$. (c, Schematic of the superconducting order parameter profile across a conventional normal-superconducting junction. (d) The ZBC profile from the normalized *dI/dV* ($V_b$=0 V) (blue symbols) from the bottom boundary region of the island. Note the ZBC is plotted with the high value (small gap) pointing down to be a measure of the superconducting order parameter. The red line is an exponential fit yielding a decay length $\xi_N$=(20±2) nm [40]. (e) The ZBC profile from the normalized *dI/dV* ($V_b$=0 V) (blue symbols), and the STM topographic height (orange symbols) across the superconducting island. The red dashed lines denote the positions of the island boundaries. (f) The ZBC profile from the normalized *dI/dV* ($V_b$=0 V) (blue symbols)

from the top boundary region of the island. The red line is an exponential fit yielding a decay length $\xi_N$=(20±2) nm [40]. Note in (d-f) the ZBC is plotted with the large values (small gap) pointing down to be a measure of the superconducting order parameter. The error bars are the standard error of the mean values.

FIG. 4. (color online) Superconducting properties a nanostructured island. (a) Temperature dependent tunneling spectra showing the transition from the superconducting to normal state measured in the center of the island of the island in Fig. 1(c). (b) The superconducting gap, $\Delta$ (symbols), as a function of temperature determined from fitting the spectra in (a) to BCS theory. The uncertainties are smaller than the symbol size. The solid line is a fit to BCS theory yielding $\Delta_0$=(221±3) μeV and $T_C$=(1.37±0.02) K [40]. (c) Magnetic field dependent tunneling spectra showing the transition to the normal state. (d) The superconducting gap, $\Delta$ (symbols), as a function of magnetic field determined from fitting the spectra in (d) to BCS theory. The uncertainties are smaller than the symbol size. From the absence of a superconductor gap, the upper critical field is estimated to be: center $B_{C2}$≈2.25 T.

FIG. 5. (color online)Sequential addition of vortices in a nanostructured superconducting island. (a) STM topographic image, 200 nm x 200 nm, of the superconducting island. (b-p) Corresponding Fermi-level $dI/dV(V_b=0)$ maps showing superconducting (red) and normal (blue) regions of the island as vortices (small blue disks) sequentially populate the nanostructure as the magnetic field is increased. The vortices are seen to distribute along the perimeter of the island and avoid the center. Image size for (b-m) is 200 nm x 200 nm. Maps (n-p) are zoomed in, 100 nm x 100 nm, to focus on the center region at higher fields.

FIG. 6. (color online)Vorticity vs normalized magnetic flux for a nanostructured superconducting island. The number of vortices in the superconducting island from Fig. 5 (blue symbols), $L$, is plotted vs the normalized magnetic flux (bottom axis) and magnetic field (top axis). The normalized magnetic flux, $\Phi_d / \Phi_0 = B \pi (D/2)^2$, where $D$ is the diameter of the island, and $\Phi_0$ is the flux quantum. The solid line represents the condition $L = \Phi_d / \Phi_0$. The blue solid lines are a linear fit from $L$=0 to 4, and $L$=5 to 8. The change in slope shows the rate of vortex addition to the island decreases by a factor of two after four vortices are added to the island. The initial slope corresponds to one vortex added for an increment of magnetic field of 0.0625 T, changing to a field increase of 0.125 T per added vortex after $L$=4. The uncertainty in the normalized flux is derived from the uncertainty in the island area, which was determined from the STM topographic profile in Fig. 3(e) [57].

FIG. A1. (color online) Optical alignment of STM probe tip onto the cleaved YPtBi crystal.

FIG. A2. (color online) STM imaging of the cleaved YPtBi surface. (a) Large area, 50 nm x 50 nm, STM topographic image of the cleaved surface. Height scale is 1.7 nm from dark to bright. (b) Zoomed in STM image, 20 nm x 20 nm, from the blue boxed region in (a). The inset shows an atomic resolution image, 5 nm x 5 nm, of several of the clusters, which make up the cleaved surface. The height scale is 1.1 nm for (b), and 0.3 nm for the inset.

FIG. A3. (color online) Al thin film tunneling spectrum. (blue circles) Tunneling $dI/dV$ spectrum from a 100 nm thick Al film measured at ≈10 mK. (red line) Modified BCS Maki calculation with parameters $\Delta$=(178±0.2) μV, $\zeta$=0.019±0.001, and $T_{eff}$=(160±5) mK [40].

FIG. A4. Sample preparation for TEM measurements. (a) Pt markers written near nanostructures for reference. (b) Carbon and tungsten deposited to protect surface during ion milling. (c) Wedge shaped sample is cut out to minimize disturbance to nanostructures. (d) Lift-out of sample. (e) Mounting and thinning of sample on TEM holder. (f) cross-sectional TEM sample.

FIG. A5. (color online) TEM measurements of a YPtBi nanostructure. (a) Annular dark-field STEM cross section image of one of the superconducting nanostructures shown in Fig. 1(d). The white arrow indicates the boundary between the amorphous layer and underlying crystal. (b) Cross sectional image from the red boxed region in (a), showing the underlying crystal structure of YPtBi and the amorphous overlayer. (c) Cross sectional image from the blue boxed region in (a), inside the nanostructure. The image shows small crystallites, which give rise to the peaks in the Fourier transform image in (d).

FIG. A6. (color online) Electrical characterization of bulk YPtBi crystals. (a) Temperature dependent resistance shows the bulk superconductivity transition. (b) Upper critical field versus temperature. The upper critical field was defined at 90 % RN, where RN is the normal state resistance. Error bars indicate field step size. (c) Temperature dependent resistance under different magnetic fields. (d) Magnetoresistance at different temperatures.

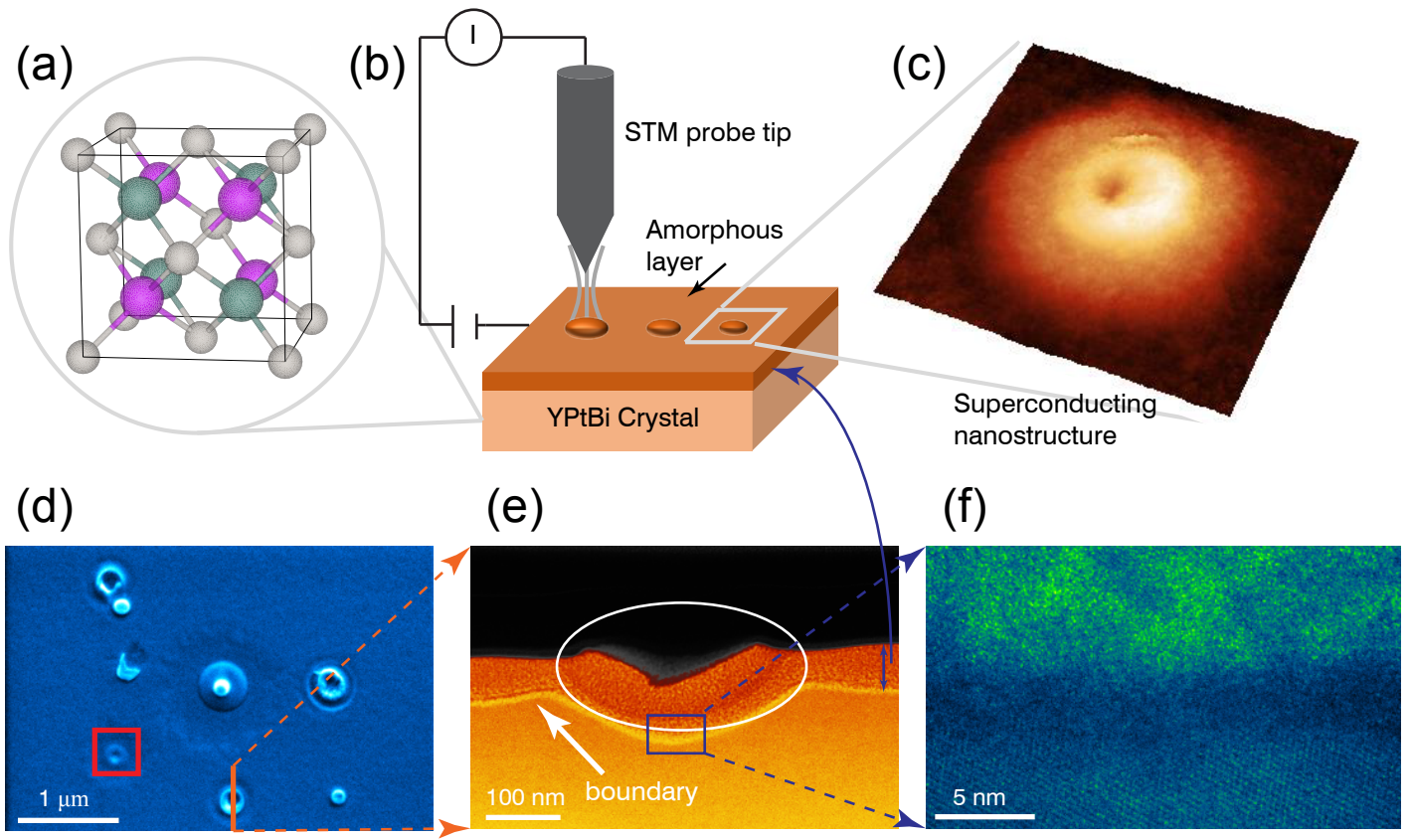

FIGURE 1.

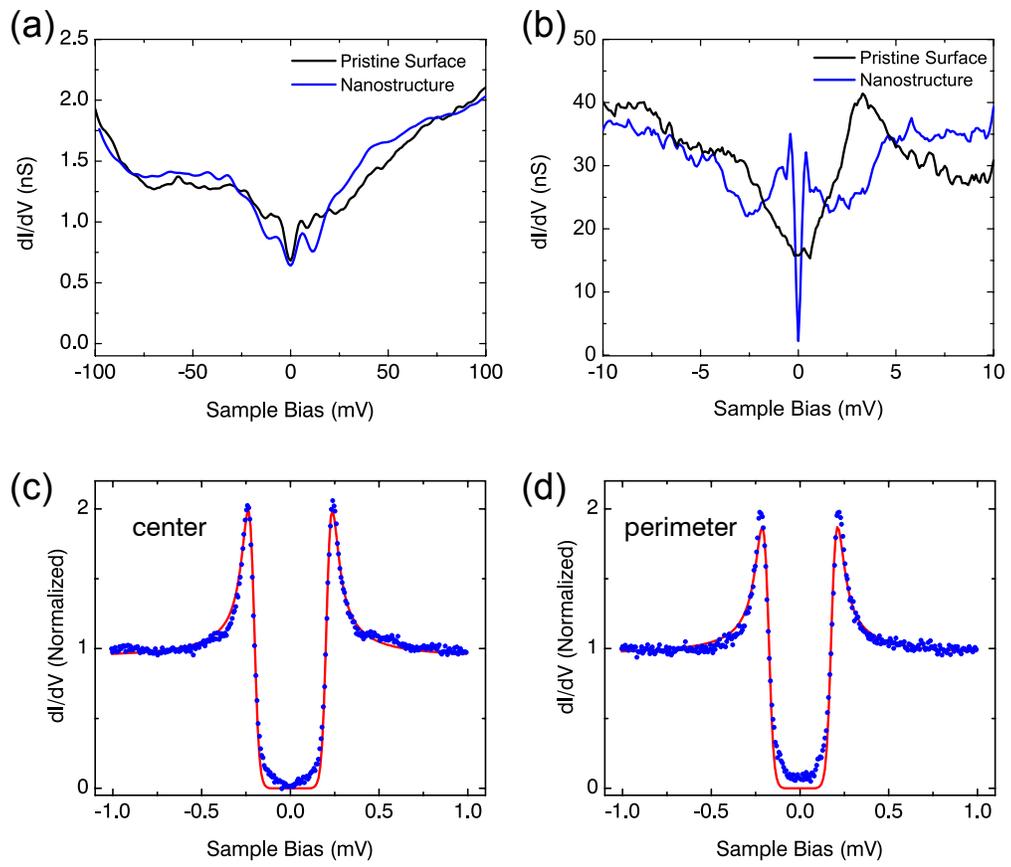

FIGURE 2.

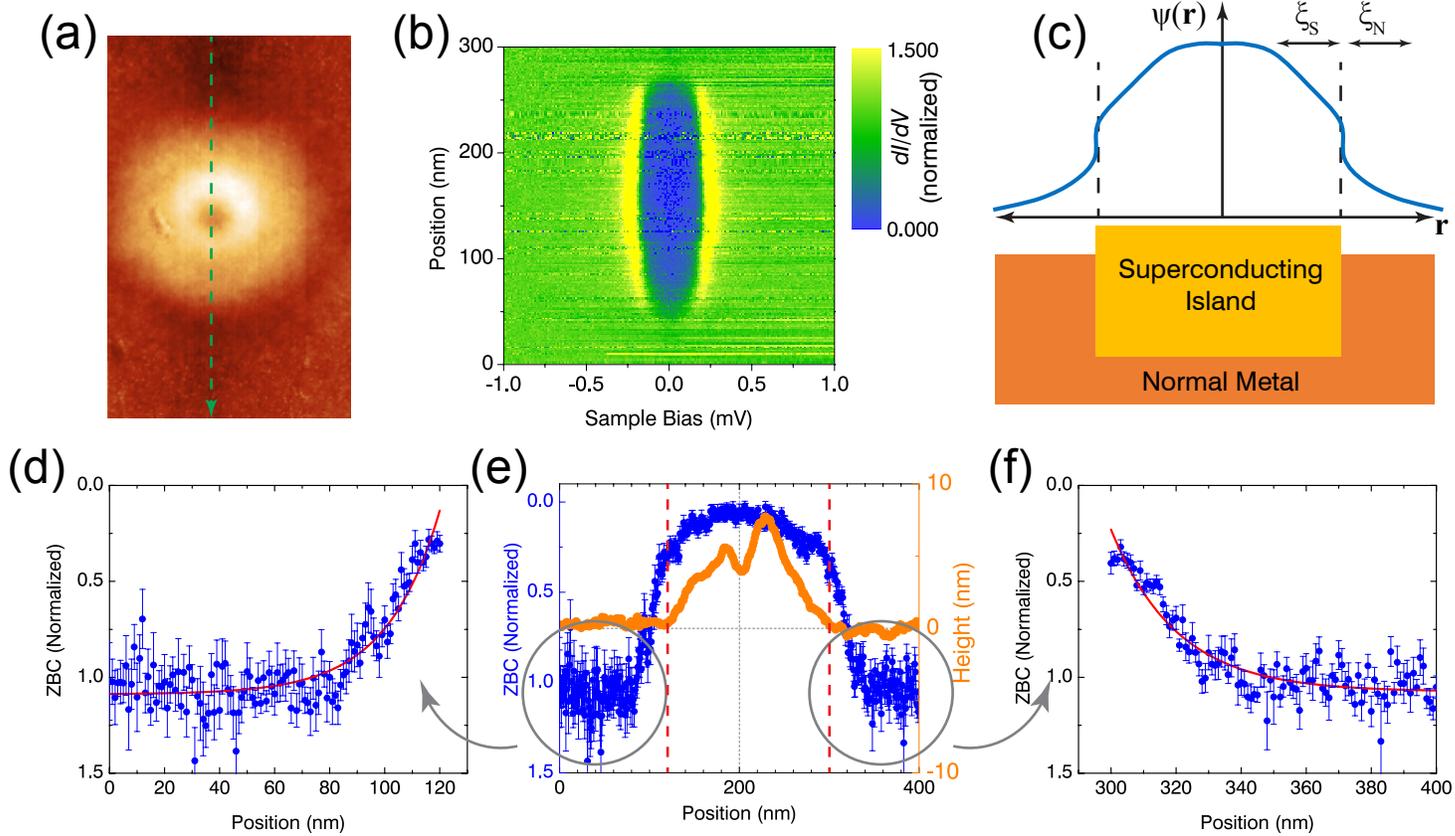

FIGURE 3.

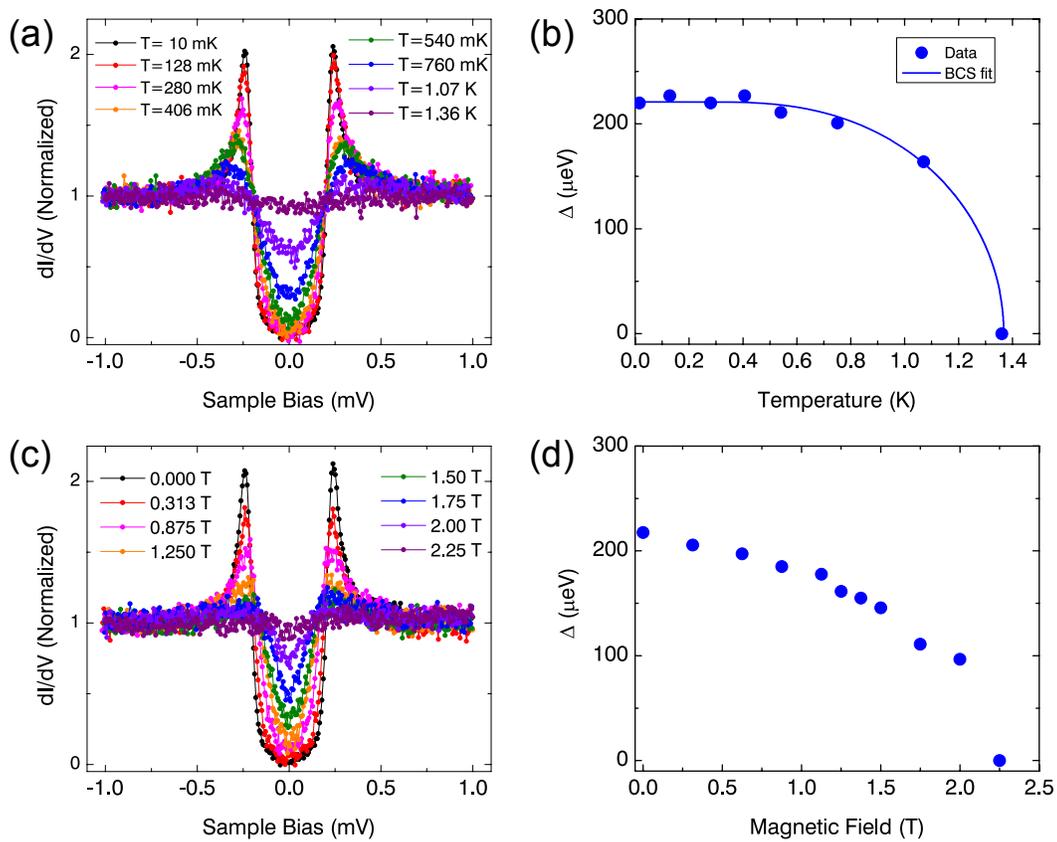

FIGURE 4

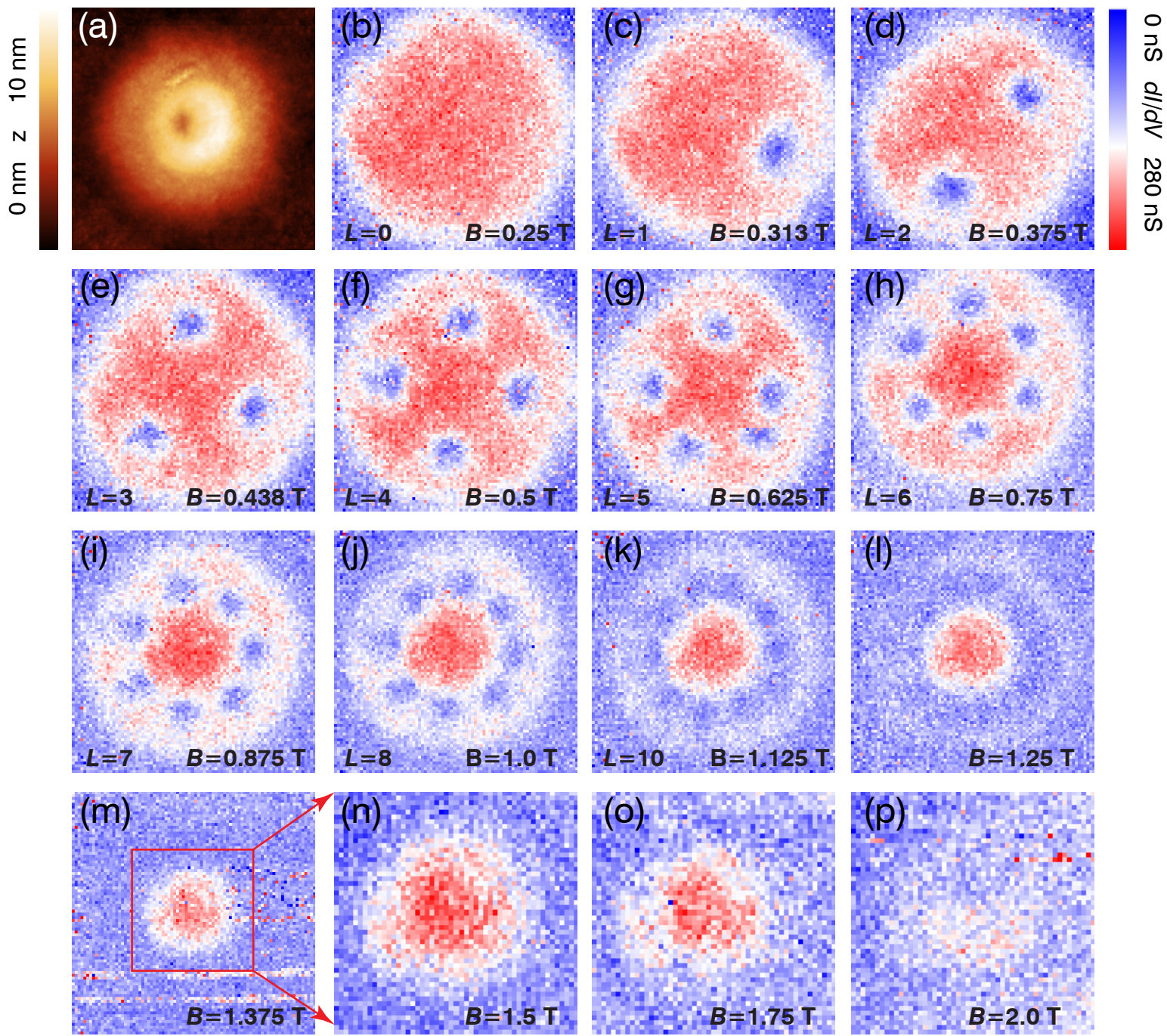

FIGURE 5.

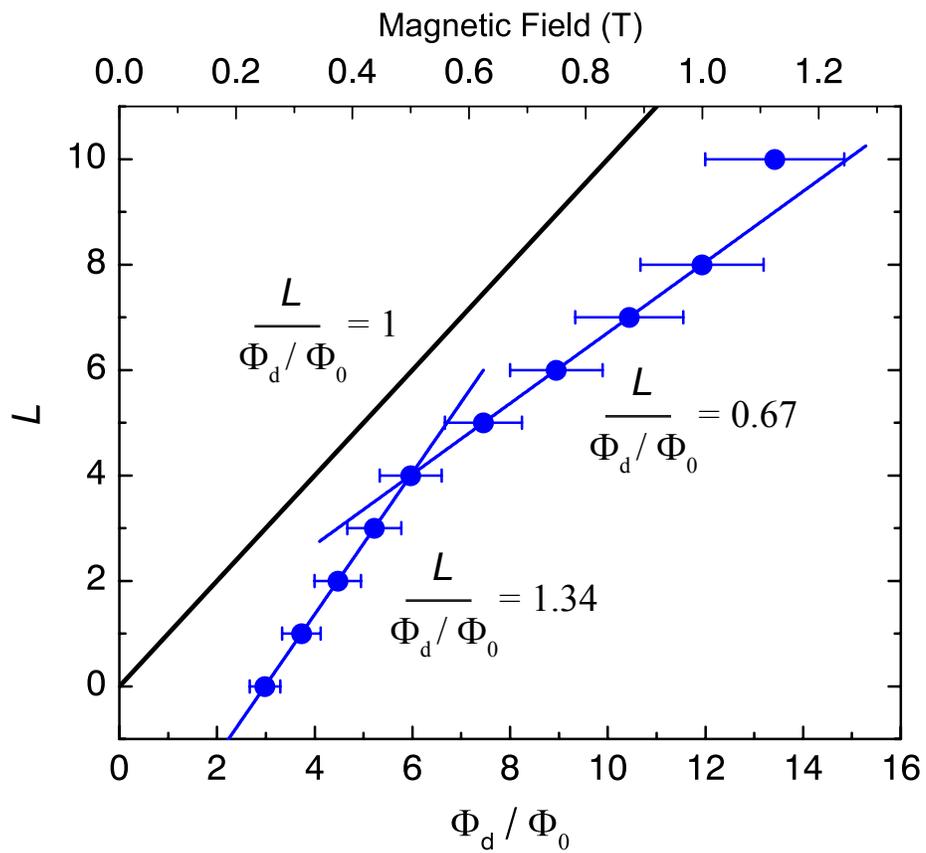

FIGURE 6.

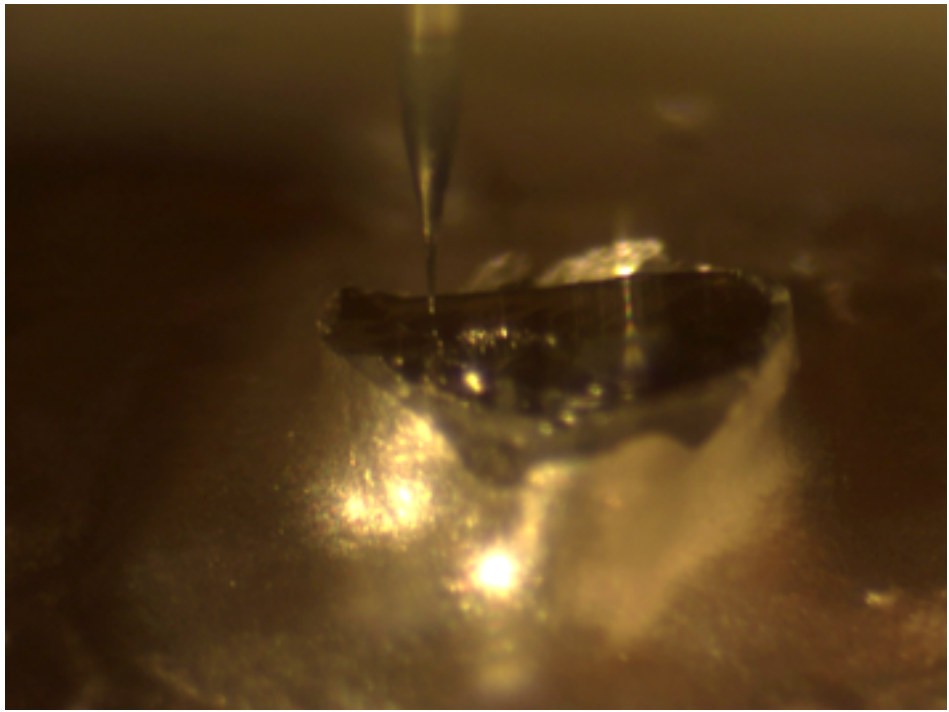

FIGURE A1.

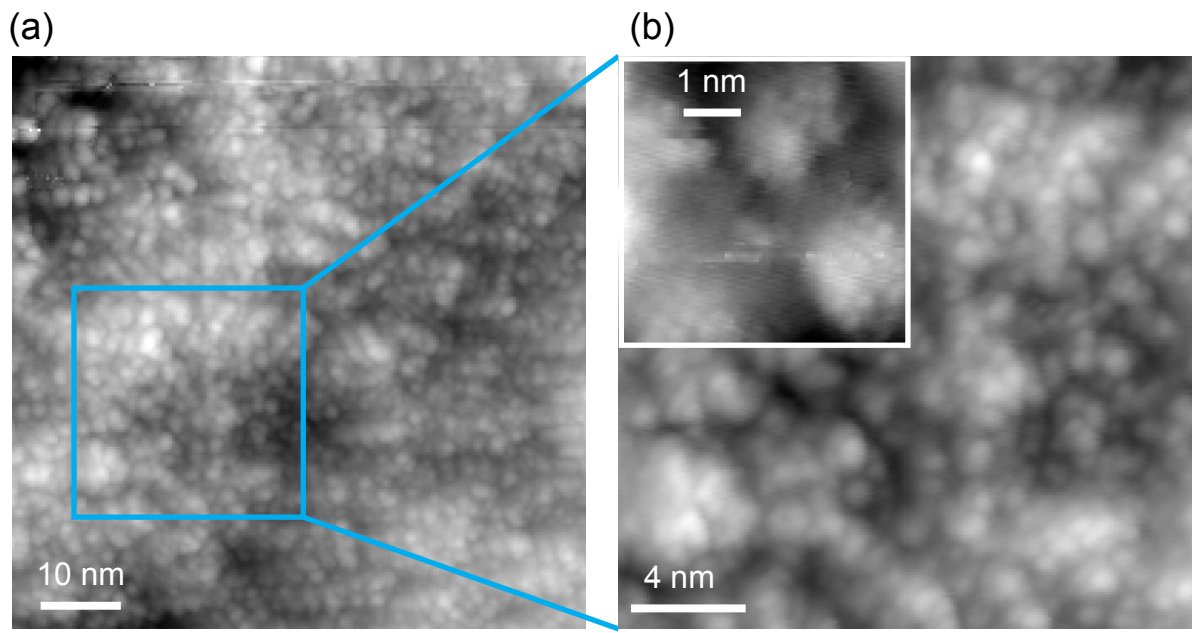

FIGURE A2.

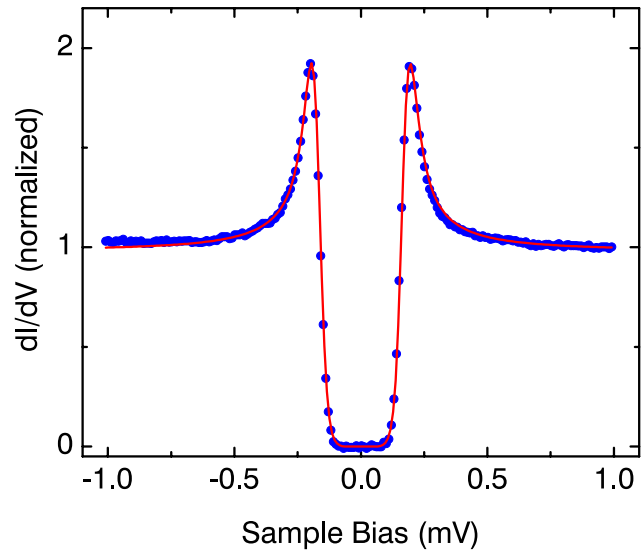

FIGURE A3.

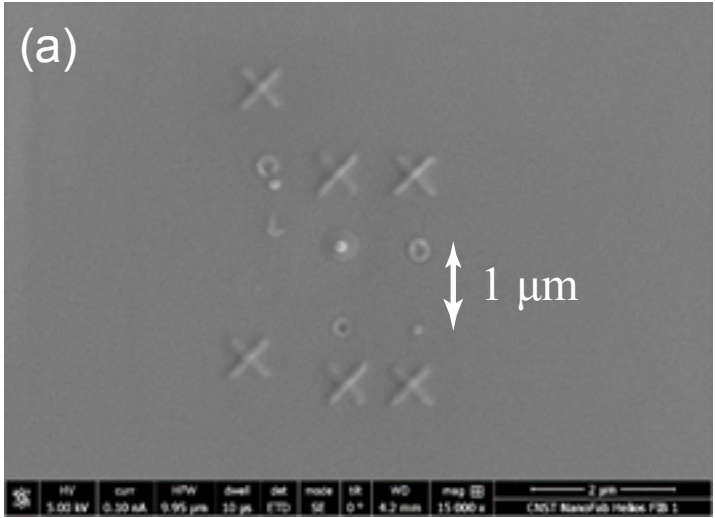
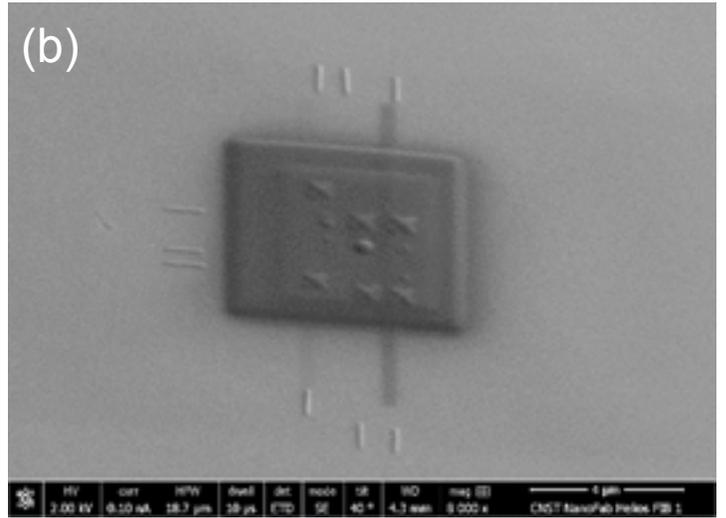
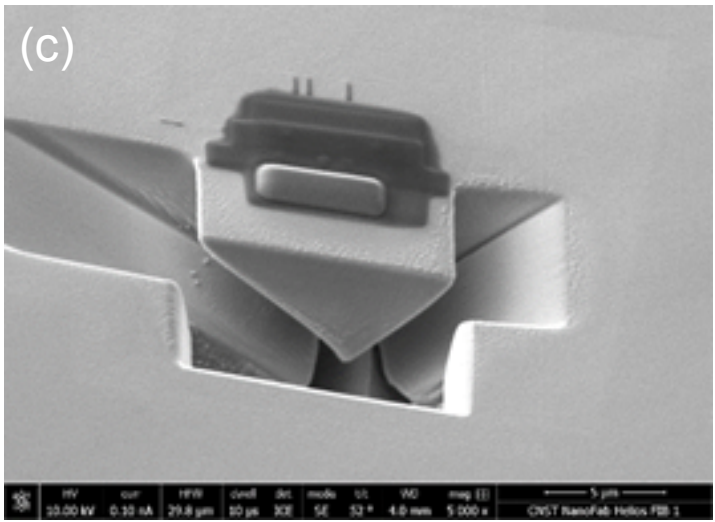
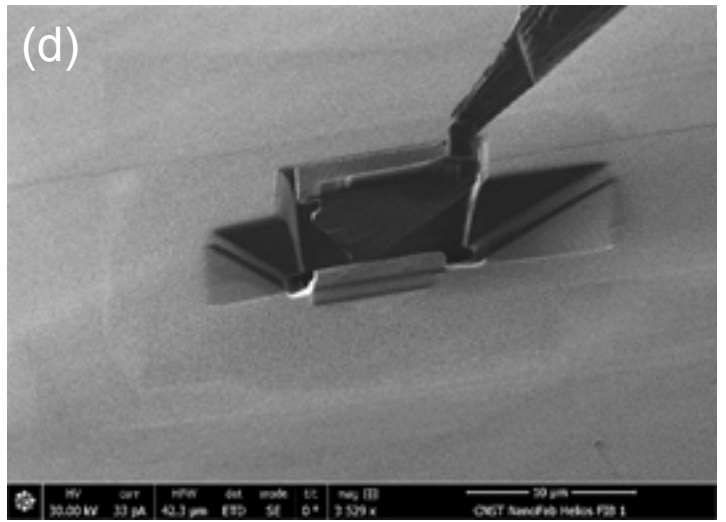
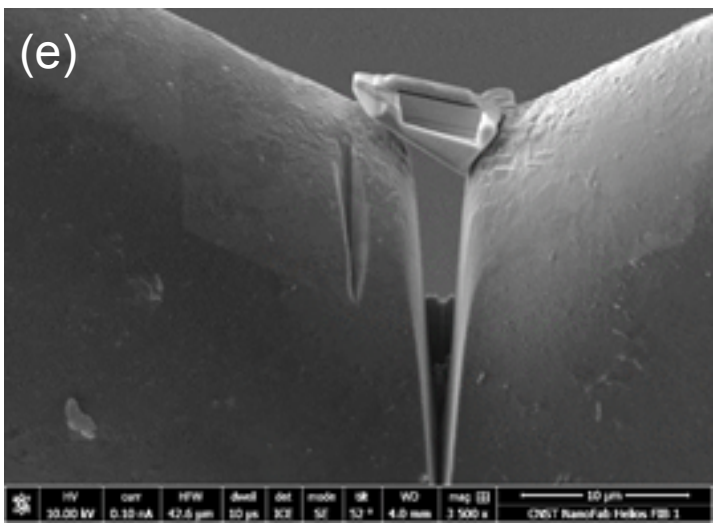
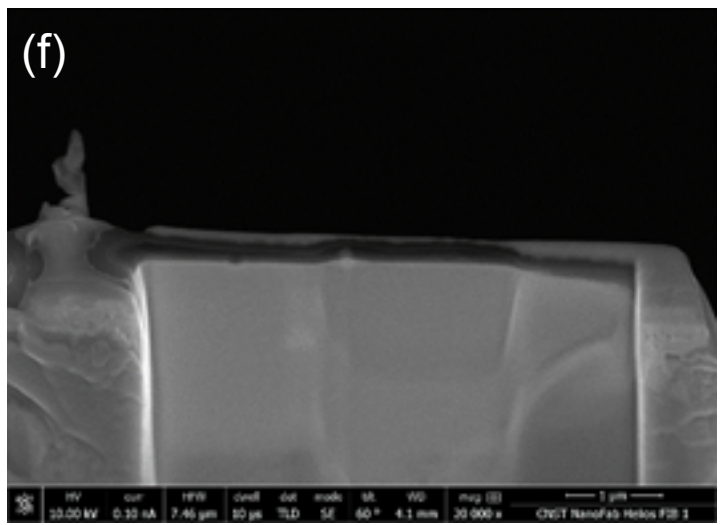

FIGURE A4.

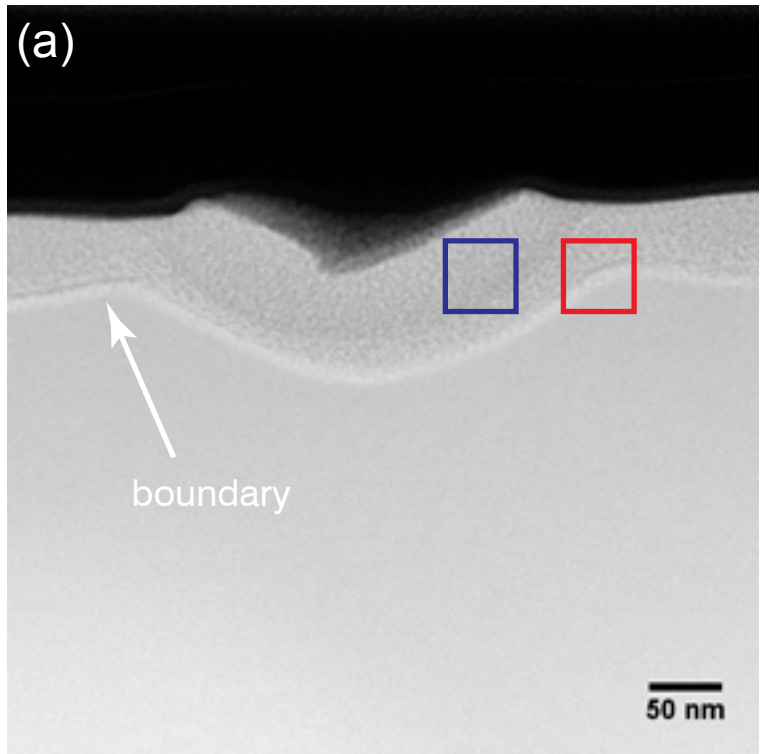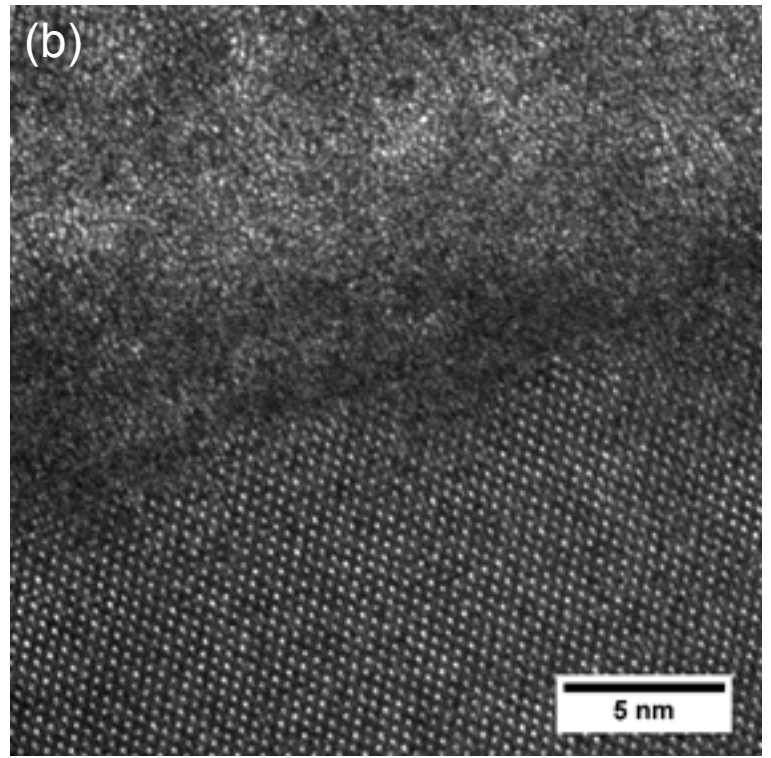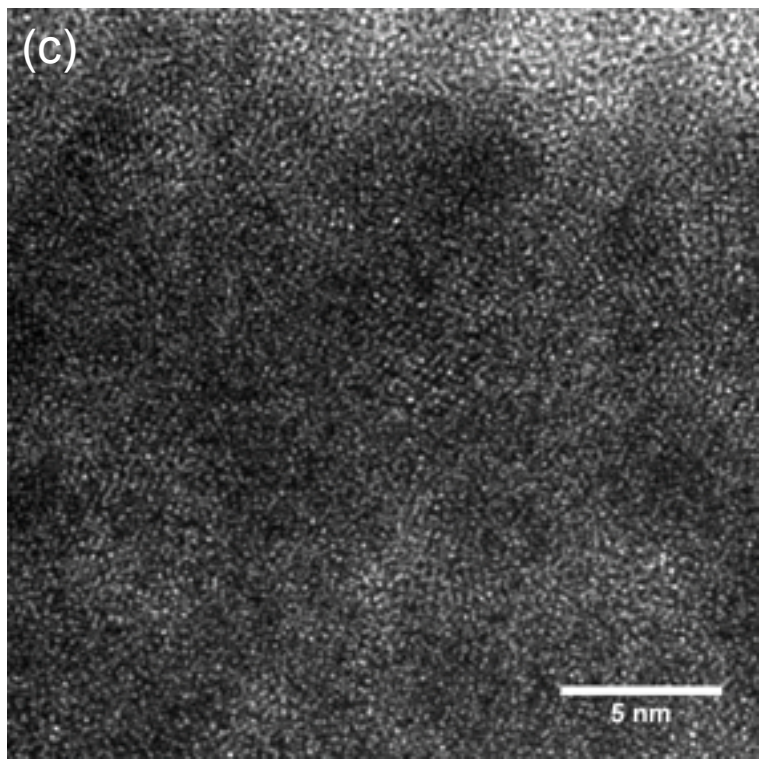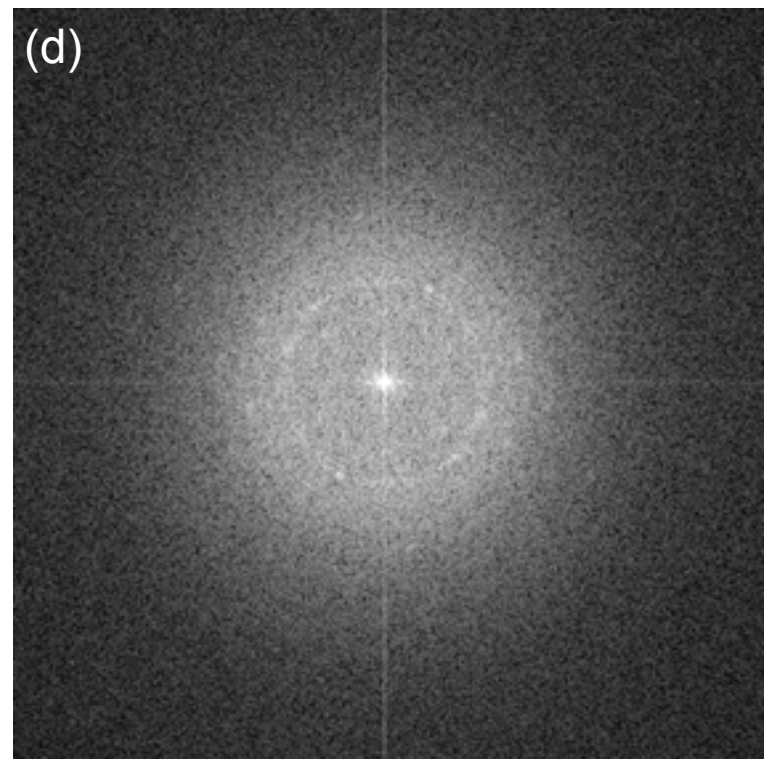

FIGURE A5.

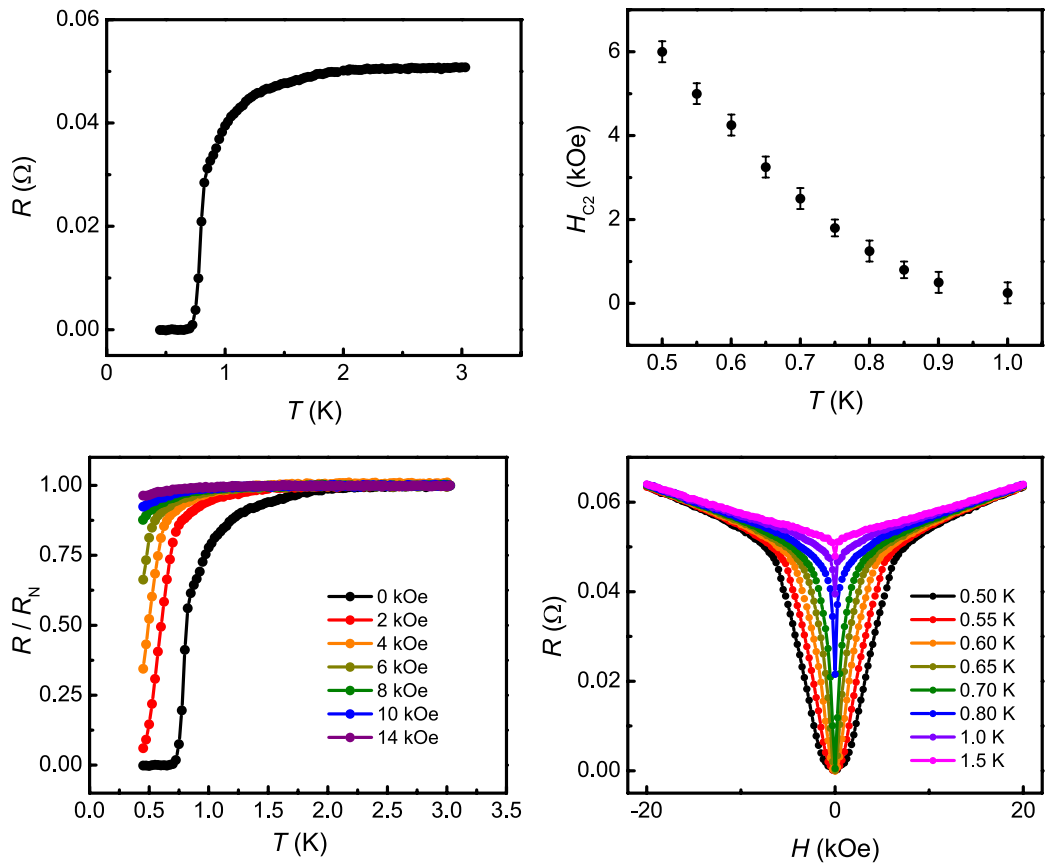

FIGURE A6.